\pdfoutput=1

\documentclass{llncs}
\usepackage{times}
\usepackage[english]{babel}
\usepackage{booktabs}
\usepackage{algorithm}
\usepackage{algpseudocode}

\usepackage{epsf,epsfig}
\usepackage{url,xspace}
\usepackage{multirow}
\usepackage{tikz}
\usetikzlibrary{calc, positioning, arrows, backgrounds}
\usepackage{pgf}
\usepackage{mathptmx}
\usepackage[T1]{fontenc}
\usepackage[utf8]{inputenc}
\usepackage{csquotes}
\usepackage{textcomp}
\usepackage{wrapfig}

\newcommand{\gb}[1]{{\mbox{$#1$~GiB}}}
\newcommand{\mb}[1]{{\mbox{$#1$~MiB}}}

\newcommand{\D}{\hphantom{0}}

\newcommand{\var}[1]{\mbox{\em #1\/}}

\newcommand{\BigO}[1]{\ensuremath{O\bigl(#1\bigr)}}

\newcommand{\myparagraph}[1]{\paragraph*{\normalsize\it#1.}}

\newcommand{\TEXT}[0]{\ensuremath{\mathsf{T}}}
\newcommand{\PATT}[0]{\ensuremath{\mathsf{P}}}
\newcommand{\STR}[0]{\ensuremath{\mathsf{STR}}}

\newcommand{\CSA}[0]{\ensuremath{\mathsf{CSA}}}
\newcommand{\SUFF}[0]{\ensuremath{\mathsf{SA}}}
\newcommand{\BWT}[0]{\ensuremath{\mathsf{BWT}}}

\newcommand{\spep}[0]{\ensuremath(\var{sp}, \var{ep})}
\newcommand{\nocc}[0]{\ensuremath{\var{nocc}}}

\newcommand{\Rank}[3]{\ensuremath{\mathsf{Rank}\bigl(#1,#2,#3\bigr)}}
\newcommand{\Rankop}[0]{\ensuremath{\mathsf{Rank}}}
\newcommand{\Selectop}[0]{\ensuremath{\mathsf{Select}}}
\newcommand{\GEQ}[2]{\ensuremath{\mathsf{GEQ}\bigl(#1,#2\bigr)}}
\newcommand{\GEQop}[0]{\ensuremath{\mathsf{GEQ}}}

\usepackage{xcolor}

\usepackage[activate={true,nocompatibility},final,tracking=true,kerning=true,spacing=true,factor=1100,stretch=10,shrink=10]{microtype}

\usepackage[colorlinks=false, pdfborder={0 0 0}]{hyperref}
\usepackage{enumitem}

\usepackage[
    backend=bibtex,
    abbreviate=true,
    style=numeric,
    firstinits=true,
    sortlocale=en_AU,
    natbib=true,
    url=false,
    doi=false,
    eprint=false,
    maxnames=99,
    maxcitenames=2,
    isbn=false
]{biblatex}
\begin{document}

\title{
CSA++:
Fast Pattern Search for Large Alphabets
}

\author{
Simon Gog\inst{1,2}
\and Alistair Moffat\inst{1}
\and Matthias Petri\inst{1}
}

\institute{
Dept. Computing and Information Systems,
The University of Melbourne, Australia
\and
Inst. Theoretical Informatics,
Karlsruhe Institute of Technology, Germany
}

\maketitle

\begin{abstract}
Indexed pattern search in text has been studied for many decades.
For small alphabets, the FM-Index provides unmatched performance, in
terms of both space required and search speed.
For large alphabets -- for example, when the tokens are words -- the
situation is more complex, and FM-Index representations are compact,
but potentially slow.
In this paper we apply recent innovations from the field of inverted
indexing and document retrieval to compressed pattern search,
including for alphabets into the millions.
Commencing with the practical compressed suffix array structure
developed by Sadakane, we show that the Elias-Fano code-based
approach to document indexing can be adapted to provide new tradeoff
options in indexed pattern search, and offers significantly faster
pattern processing compared to previous implementations, as well as
reduced space requirements.
We report a detailed experimental evaluation that demonstrates the
relative advantages of the new approach, using the standard
Pizza\&Chili methodology and files, as well as applied use-cases
derived from large-scale data compression, and from natural language
processing.
For large alphabets, the new structure gives rise to space
requirements that are close to those of the most highly-compressed
FM-Index variants, in conjunction with unparalleled search throughput
rates.

\keywords{String search, pattern matching, suffix array,
Burrows-Wheeler transform, succinct data structure, experimental
evaluation.}
\end{abstract}

\section{Introduction and Background}
\label{sec-introduction}

We study a well-known problem:
given a static text $\TEXT[0,n-2]$ over an alphabet $\Sigma$ of size
$\sigma$ followed by a symbol $\TEXT[n-1]=\$$, with
$\$\not\in\Sigma$, preprocess $\TEXT$ so that a sequence of patterns
$\PATT[0,m-1]$, also over $\Sigma$, can be efficiently searched for,
with the purpose of each search being to identify the number of
occurrences $\nocc$ of $\PATT$ in~$\TEXT$.
A variety of options exist for this problem, providing different
trade-offs between construction cost, memory space required during
pattern search operations, and search cost, both asymptotically and
in practical terms.
Example structures include the {\emph{suffix tree}}
{\cite{acfgm-cacm16,w-swat73}} and {\emph{suffix array}}
{\cite{mm93siamjc}}.
The suffix array of $\TEXT$, denoted {\SUFF}, requires $\BigO{n\log
n}$ bits of space in addition to the $\BigO{n\log\sigma}$ bits
occupied by $\TEXT$, and uses that space to store the offsets
$\SUFF[0,n-1]$ of all $n$ suffixes of $\TEXT$ (denoted as
$\TEXT[i,]$) in lexicographic order such that
$\TEXT[\SUFF[i],]<\TEXT[\SUFF[i+1],]$ for $i\in [0,n-1]$.
Using {\SUFF}, the number of occurrences of $\PATT$ in $\TEXT$ can be
identified in $\BigO{m\log n}$ time, via two binary searches that
determine the range $\spep$ such that all suffixes
$\SUFF[\var{sp},\var{ep}]$ are prefixed by $\PATT$.
Thus, $\nocc=\var{ep}-\var{sp}+1$.
The search cost can be reduced to $\BigO{m+\log n}$ if
information about longest common prefixes is also available.
Storing this information for all possible intervals $\SUFF[i,j]$
occurring in the binary search process requires
$\BigO{n\log n}$ bits of additional space.

\myparagraph{Compressed Indexes}

In a {\emph{compressed suffix array}}, or {\CSA},
the space required is proportional to the compressed size of $\TEXT$.
{\textcite{s-jalg03}} (see also {\textcite{gv-stoc00}}) describes a
{\CSA} based on the observation that the function
$\psi[i]={\SUFF}^{-1}[(\SUFF[i]+1)\mbox{~mod~}n]$
consists of $\sigma$ increasing sequences (or {\emph{segments}}) of
integers, and that each of those segments is likely to be
compressible, yielding a space usage of
$nH_k(\TEXT) + \BigO{n \log \log \sigma}$
bits~\cite{nm-csurv07}, where $H_k$ denotes to the order-$k$
empirical entropy of $\TEXT$.
Occurrences of $\PATT$ are located by performing a {\emph{backward
search}}
to find the
range $\SUFF[\var{sp}_i,\var{ep}_i]$ matching each suffix $\PATT[i,]$,
stopping if $\var{ep}_i<\var{sp}_i$, or if all of $\PATT$ has been
processed.

An alternative compressed indexed is due to {\textcite{fm00focs}},
and is based on the Burrows Wheeler Transform (\BWT), defined as
$\BWT[i]=\TEXT[\SUFF[i]-1 \bmod n]$.
In an {\emph{FM-Index}} the $\BWT$ is generally encoded using a
{\emph{wavelet tree}} {\cite{ggv03soda}}, and accessed using
$\Rank{\BWT}{i}{c}$, which returns the number of times symbol $c$
occurs in the prefix $\BWT[0,i-1]$.
Again, $\PATT$ is processed in reverse order.
Suppose $\SUFF[\var{sp}_i,\var{ep}_i]$ refers to the range in $\SUFF$
prefixed by $\PATT[i,]$, and that $\PATT[i-1]=c$.
An array $C$ of $\sigma \log n$ bits stores the number of symbols $c$
in $\TEXT$ smaller than $c$; using it,
$\var{sp}_{i-1} = C[c] + \Rank{\BWT}{\var{sp}_i}{c}$
and
$\var{ep}_{i-1} = C[c] + \Rank{\BWT}{\var{ep}_i+1}{c}-1$
can be computed.
Overall, $\SUFF[\var{sp},\var{ep}]$ is identified using $2m$
$\Rankop$ operations on the {\BWT}; and when stored using a wavelet
tree, $\BigO{m \log \sigma}$ time.
For more information about these structures, and the time/space
trade-offs that they allow, see {\textcite{nm-csurv07}} and
{\textcite{fgnv-jea08}}.

\myparagraph{In Practice}

Implementations of the {\CSA} and the FM-Index have been developed
and measured using a range of data.
When $\sigma$ is small -- for example, $\sigma=4$ for DNA, and
$\sigma \approx 100$ for plain ASCII text -- both provide fast
pattern search based on compact memory footprints, usually requiring
half or less of the space initially occupied by $\TEXT$, depending on
a range of secondary structures and parameter choices
{\cite{gnp-jda15,hcvn-dcc14}}, and with the FM-Index typically
requiring less space that the {\CSA}.
But when $\sigma$ is large -- for example, when the
alphabet is words in a natural language and $\sigma\approx 10^6$ or
greater -- the situation is more complex.
In particular, the $\BigO{\log\sigma}$ factor associated with the
FM-Index's wavelet tree is a count of random accesses (as distinct
from cache-friendly accesses) and means that search costs increase
with alphabet size, negating its space advantage.
In contrast, standard {\CSA} implementations are relatively
unaffected by $\sigma$, but each backward search step in a {\CSA} has
a dependency on $\log n_c$, where $n_c$ is the frequency in $\TEXT$
of the current symbol $c=\PATT[i]$.
Hence, if $\sigma$ is fixed and does not grow with $n$, {\CSA}
pattern match times will grow as $\TEXT$ becomes longer.

\myparagraph{Our Contribution}

We introduce several improvements to the {\CSA} index:
\begin{itemize}
\item
We adapt and extend the uniform partitioned Elias-Fano (UEF) code
of {\citet{ov14sigir}} to the storage of the $\psi$ function,
allowing faster backwards search compared to previous
implementations;
\item
We add a fourth UEF block type compared to {\citeauthor{ov14sigir}},
and include the option of coding sections of the $\psi$ function in a runlength mode;
\item
We describe a way of segregating the short segments in $\psi$,
allowing improved compression when
$\sigma$ is large and many of the symbols in $\Sigma$ are rare;
\item
We carry out detailed ``at scale'' experiments, including
both synthetic query streams and logs derived from
use-cases, covering all of small, medium, and large alphabets.
\end{itemize}
The result is a pattern search index that we refer to as ``CSA++''.
It represents a significant shift in the previous relativities
between compressed index structures; and, for large alphabets in
particular, gives rise to space needs close to those of the most
highly-compressed FM-Index variants, with unparalleled
search throughput rates.

\section{Storing Integer Lists}
\label{sec-csa}

\myparagraph{Operations Required}

The function $\psi[i]=\SUFF^{-1}[(\SUFF[i]+1)\mbox{~mod~}n]$ is a
critical -- and costly -- component of a {\CSA}.
It can be thought of as consisting of a concatenation of $\sigma$
{\emph{segments}}, the $c$\,th of which is a sorted list of the
locations in $\BWT$ at which the $c$\,th symbol in $\Sigma$ appear.
That is, each segment of $\psi$ can be interpreted as a
{\emph{postings list}} of occurrences of symbol~$c$.
The key operation required to enable backwards search is that of
$\GEQ{c}{\var{pos}}$, which returns the smallest position
$\var{pos}'$ such that $\psi[\var{pos}']$ is in the $c$\,th segment,
and such that $\psi[\var{pos}']\ge\var{pos}$.
Starting with $\var{sp}=0$ and $\var{ep}=n-1$, the $\spep$ bounds are
narrowed via a sequence of $m$ pairs of $\var{sp}=\GEQ{c}{\var{sp}}$
and $\var{ep}=\GEQ{c}{\var{ep}+1}-1$ operations, as $c$ takes on
values from $\PATT[m-1]$ through to $\PATT[0]$.
The equivalence of the {\CSA} and FM-Index search processes can be
seen by noting that
$\GEQ{c}{\var{pos}}=C[c]+\Rank{\BWT}{\var{pos}}{c}$, and that all of
the $\spep$ pairs computed are identical between the two.
Note also that, by construction, symbol occurrences in the {\BWT}
string are likely to appear in clusters, and hence $\psi$ is likely
to contain runs of consecutive or near-consecutive integers,
separated by large intervals, and to contain at most $\sigma$
``disruption'' points at which $\psi[i]>\psi[i+1]$.

\myparagraph{Integer Codes}

One common way of storing postings lists is to compute {\emph{gaps}},
or {\emph{differences}}, and then store them using a suitable code
for integers; clusters in $\BWT$ then gives rise to runs of small or
unit gaps in $\psi$.
A range of integer codes have been developed for this type of
distribution, including Elias $\gamma$ and $\delta$ codes, Rice and
Golomb codes, and the Binary Interpolative Code (see
{\citet[Chapter~3]{mt02caca}} for descriptions).
Several of these have been used in previous {\CSA} implementations
{\citep{s-jalg03}}.

There has been recent interest in {\emph{Elias-Fano codes}} (EF
codes) for postings list compression, a result of work by
{\citet{vigna13wsdm}} (see also {\citet{am98sigir}} for earlier
application, and {\citet{gnp-jda15}} for preliminary experimentation
with compressed suffix arrays).
Given a non-decreasing set of $k$ integers in the range $0\ldots
2^U-1$ for some universe size $2^U$, a parameter $\ell$ is selected,
and each integer is split into a high part (the most significant
$U-\ell$ bits) and a low part (the $\ell$ low-order bits).
Groups are formed for values that have the same high parts.
A code for the block of $k$ values is then constructed by
representing the size of each of the $2^{U-\ell}$ possible groups in
unary, followed by the concatenation in order of the $k$ low parts.
For example, if $U=4$ and $k=3$, the sequence $[6, 7, 10]$ (that is,
$[0110, 0111, 1010]$ in binary) would be coded using $\ell=2$, and
split into high parts, $[01,01,10]$, coded as group sizes in unary as
{\tt{0:110:10:0}}; and into low parts coded in binary,
{\tt{10:11:10}}, where the ``{\tt{:}}''s are purely indicative, and
do not appear in the output.
The EF code achieves representations close to the combinatorial
minimum if $\ell=\lfloor\log_2(2^U/k)\rfloor$; moreover, the length
of the coded block is easily computed: $k+2^{U-\ell}$ bits are
required for the high/unary parts, and $k\cdot\ell$ bits for the
low/binary parts.

One useful aspect of the EF code is that the unary parts can be
searched via {\Selectop} operations over their ``{\tt{0}}'' bits, and
then the number of binary parts through until that point computed.
For example, in the unary sequence shown above, any elements from the
underlying sequence in the range $8\ldots11$ must fall in the third
bucket, and $\Selectop_0(2)-2=4-2=2$ indicates that there are in
total two binary parts contained within the first two buckets, and
hence that the binary parts associated with the third bucket (if any)
must commence from the third element of the low/binary part.
On average there is $\BigO{1}$ item per bucket, and linear search can
be used to scan them; if a worst-case bound is required, binary
search can be used if there are more than $\log_2 n$ ``{\tt{1}}''
bits between the relevant pair of consecutive ``{\tt{0}}'' bits, and
linear search employed otherwise.

Another feature of EF codes is that in the binary part all components
are of the same bit-length~$\ell$, meaning that there are no
dependencies that would hinder vectorized processing and
loop-unrolling techniques and prevent them from achieving their full
potential.
This is not the case with, for example Elias $\delta$ codes, which
are based upon gaps and are also of variable length, and hence must
be decoded sequentially.

\myparagraph{Partitioned Elias-Fano Codes}

The term occurrences in long postings lists tend to be clustered, a
pattern that has been used as the basis for a range of improved index
compression techniques {\citep{mt02caca}}.
{\citet{ov14sigir}} demonstrated that EF codes could capture much of
this effect if postings lists were broken into blocks of $k$ values,
and then the document identifiers in each block mapped to the range
$0\ldots 2^U-1$ for some suitable per-block choice of $U$.
{\citeauthor{ov14sigir}} further observed that in some cases EF codes
are less efficient than other options, and that it was helpful for
blocks to be coded in one of three distinct modes: (i) those
consisting of an ascending run of $k$ consecutive document
identifiers, in which case no further code bits are required at all
(NIL blocks); (ii) those where the document identifiers are
sufficiently clustered (but not consecutive) that a $2^U$-bit vector
is the most economical approach (BV blocks); and (iii) those that are
best represented using EF codes, taking $2^{U-\ell} + k\cdot(1+\ell)$
bits.
Note that the decision between these options can be made based solely
on $k$ and $U$.

The combination of fixed-$k$ blocks and
range-based code selection is referred to as {\emph{Uniform
Elias-Fano}} (UEF) coding.
{\citeauthor{ov14sigir}} also describe a mechanism for partitioning
postings lists into approximately-homogeneous variable-length blocks
in a manner that benefits EF codes that we do not employ here.

\myparagraph{Overall Structure of a {\CSA}}

With gaps in $\psi$ represented by variable-length codewords,
the ability to directly identify and then search segments
of $\psi$ is lost.
Instead, pseudo-random access is provided via a set of
{\emph{samples}}: $\psi$ is broken into fixed-length blocks; the
first $\psi$ value in each block is retained uncompressed in a
{\emph{sample index}}; and the remaining values in that block are
coded as gaps starting from that first value
{\citep{os07alenex,s-jalg03}}.
Computation of $\GEQ{c}{\var{pos}}$ then involves identification of
the region in the sample index associated with the segment for
symbol~$c$, binary search in that section of the sample index to
identify the single block that contains $\var{pos}$ or the next
$\psi$ value greater than it; and then sequential decoding of that
whole block, to reconstruct values of $\psi$ in order to determine
the exact value.
If symbol $c$ occurs $n_c$ times in $\TEXT$, and if samples are
extracted every $k$ values, then searching the sample index requires
$\BigO{\log(n_c/k)}$ time, a cost that must be balanced against the
$\BigO{k}$ cost of linear search within the block.
Small values of $k$ give faster $\GEQ{c}{\var{pos}}$ operations, but
also increase the size of the sample index, and hence the size of the
{\CSA}.

\section{Representing $\psi$}
\label{sec-newcsa}

We store the $\psi$ function of a {\CSA} using the UEF approach of
{\citeauthor{ov14sigir}}, using a blocksize of $k$ as the basis for
both the UEF code and the sample index {\citep{gnp-jda15}}.
A number of further enhancements to previous implementations are now
described.

\myparagraph{Independent Structures}

Rather than storing the whole of $\psi$ as a single entity split into
blocks, we treat each segment independently, and genuinely form an
inverted index for the symbols $c$ in $\BWT$.
The $\sigma\log n$-bit array $C$ of cumulative symbol frequencies is
retained, and hence $n_c=C[c+1]-C[c]$.
A UEF-structured postings list of $\lceil n_c/k\rceil$ blocks is then
created for symbol $c$, with its own sample index constructed from
the first (smallest) value in each of the blocks, and also
represented using an EF code, with $U'=\lceil\log_2 n\rceil$ as the
universe size for this ``top level'' structure, and $k'=\lceil
n_c/k\rceil$ the number of values to be coded within it.

One risk with this ``separate structures'' approach is that symbols
$c$ for which $n_c$ is small may incur relatively high overheads; a
mechanism for addressing this concern is presented shortly.
Another potential issue is the cost of the mapping needed to provide
access to the $c$\,th of these structures, given a symbol identifier
$c$; that process is also described in more detail later in this
section.

\myparagraph{RL Blocks}

{\citet{ov14sigir}} employ three block types, to which we add a
fourth: {\emph{run-length encoded}} blocks (RL blocks).
The NIL blocks of {\citeauthor{ov14sigir}} account for runs of $k$
consecutive $\psi$ values; but there are also many instances of
shorter runs that do not span a whole block.
In an RL block, the (strictly positive) gaps between consecutive
$\psi$ values are represented using the Elias $\delta$ code.
Any unit gaps are followed by a second $\delta$ code to indicate a
repeat counter, while non-unit gaps are left as is.
For example, $[27, 28, 29, 45, 46, 47, 48, 70, 71, 73]$ would be
represented as $[(+1,2), +16, (+1,3), +22, (+1,1), +2]$, with the
plus symbols and parentheses indicative only, and with the sampled
value $27$ held in the top-level structure.

To decide whether to apply RL mode to any given block, the space that
it would consume is found by summing the lengths of the $\delta$
codes, and comparing against the (calculated) cost of the BV and EF
alternatives.
Because $\delta$ is slower to decode than EF codes, a ``relative
advantage'' test is applied, and blocks are coded using the RL
approach only if the RL size is less than half the size of the
smaller of an equivalent BV or EF-coded block.
A flag bit at the start of each block informs the decoder which mode
is in use for that block.

\myparagraph{Low-Frequency Symbols}

When $\sigma$ is large it is likely that many symbols in $\Sigma$
have relatively low frequencies and hence notably different values in
$\psi$; and having a small number of widely-spaced values in a block
that is otherwise tightly clustered increases the cost of every
codeword in the block, because of the non-adaptive nature of the EF
code.
In the ``separate structures'' approach we are adopting, there is
also a level of per-segment overhead that is relatively expensive for
short segments.
To address this issue, we add a further option for storing the $\psi$
values for low-frequency symbols, and do not build an independent UEF
structure for them.
For example, consider a symbol $c$ of frequency $n_c=2$.
Its segment in $\psi$ is only two symbols long, and it is far more
effective to segregate those two values into two elements of a
separate array using $\lceil\log_2 n\rceil$ bits each than it is to
construct a UEF structure and the associated sample index.
In particular, if those two elements are within a larger array in
which all of the values for all symbols for which $n_c=2$ are stored,
the overhead space can be kept small.

The array $C$ has already been mentioned, it allows $n_c$ to be
computed for a symbol~$c$.
A bitvector $D$ of size $\sigma$ with $\Rankop$ support is
added, with $D[c]=1$ if symbol $c$ is being stored as a full UEF
structure, and $D[c]=0$ if $n_c\leq L$ for some threshold~$L$. We use
$D$ to map from $\Sigma$ to $\Sigma'=\{c\mid n_c\leq L\}$.
The next component required is a wavelet tree over the values $n_c$,
where $c\in\Sigma'$, to support $\Rankop$ operations and hence
determine how many symbols $c'<c$ in $\Sigma'$ have $n_{c'}=n_c$.
Finally, a set of $L$ arrays are maintained, one for each symbol
frequency between $1$ and $L$.
We suppose that $A_i$ is the $i$\,th of those arrays.
With those components available, locating the segment of $\psi$
values corresponding to symbol $c$ is carried out as follows.
First, $D[c]$ is accessed and $n_c=C[c+1]-C[c]$ is determined.
If $D[c]$ is zero, the wavelet tree is used to compute $s=|\{ c' \mid
1\le c'< c\mbox{~and~}n_{c'}=n_c\}|$, and the $n_c$ required values
of $\psi$ are at $A_{n_c}[n_c\cdot s\ldots n_c\cdot s+n_c-1]$.
On the other hand, if $D[c]=1$, then $s=\Rank{D}{c}{1}$ is computed,
and the $s$\,th of the full UEF structures is used to access the
$c$\,th segment of $\psi$.

In the experiments reported in the next section we take $L=k$, where
$k$ is the UEF block size and also the sample interval.
That is, any symbols $c$ for which $n_c\le k$ and less than one full
UEF block would be required are stored in uncompressed form as binary
values in the range $0\ldots n-1$, in contiguous sections of shared
arrays.
Note that as a further small optimization the groups of $n_c$
elements that collectively comprise each of the arrays $A_{n_c}$
could themselves be stored using EF codes when $n_c\ge2$, since the
EF-compressed length of each such group is both readily calculable
and identical.
However, given that naturally-occurring large-alphabet frequency
distributions typically have long tails of very low symbol
frequencies, the average cost of such EF codes might be close to
$\lceil\log_2 n\rceil$ bits per $\psi$ value anyway, in which case we
would expect the additional gains to be modest.
We leave detailed exploration of this idea for future work.

\myparagraph{Eliminating Double Search}

As described earlier, each symbol that is processed in $\PATT$ gives
rise to two $\GEQop$ operations over $\psi$.
It is thus tempting to compute these via two calls to the same
function.
But much of the computation between the two calls can be shared, and
it is more efficient to perform the first $\GEQop$ call to identify
$\GEQ{c}{\var{sp}}$, and then perform a finger-search from that point
to compute the equivalent of $\GEQ{c}{\var{ep}}$.

\section{Experiments}
\label{sec-experiments}

\myparagraph{Methodology and Implementation}

The baselines and CSA++ are written in {\tt{C++14}} on top of the
SDSL library {\citep{gbmp2014sea}} and compiled with optimizations
using {\tt{gcc~5.2.1}}.\footnote{To ensure the reproducibility of our
results, our complete experimental setup, including data files, is
available at {\url{github.com/mpetri/benchmark-suffix-array/}}.}
We also make use of {\citeauthor{s-jalg03}}'s source code as a
further reference point {\citep{s-jalg03}}.
The experimental results were generated using a Intel Xeon E5640 CPU
using {\gb{144}} RAM.
All timings reported are averaged over five runs; the variance was
low and all measurements lie within approximately $10\%$ of each reported
value.
All space usages reported are those of the serialized data structures
on disk.

\myparagraph{Data Sets, Queries and Test Environment}

Our experiments make use of texts $\TEXT$ from two different sources:
four $\mb{200}$ files drawn from the Pizza\&Chili
corpus\footnote{See
{\url{http://pizzachili.dcc.uchile.cl/texts.html}}.}, selected to
illustrate a range of alphabet sizes $\sigma$; plus two $\gb{2}$
files of natural language text, one in German, and one in Spanish.
The latter were extracted from a sentence-parsed prefix of the German
and Spanish sections of the CommonCrawl\footnote{See
{\url{http://data.statmt.org/ngrams/deduped/}}}.
The four $\mb{200}$ Pizza\&Chili files are treated as byte streams,
with $\sigma\le 256$ in all cases; the two larger files are parsed in
to word tokens, and then those tokens mapped to integers.
There were $\sigma=5{,}039{,}965$ distinct words (integers) in the
German-language file, and $\sigma=2{,}956{,}209$ distinct words in
the Spanish-language file.

The primary query streams applied to these files were generated by
randomly selecting $50{,}000$ locations in $\TEXT$ and extracting
$m=20$-character strings for the Pizza\&Chili files, and extracting
$m=4$-symbol/word strings for the two natural language files.
This follows the methodology adopted by other similar experimentation
carried out in the past.
As secondary query streams, we also make use of the strings generated
by two specific use-cases, described later in this section, in part
as a response to the concerns explored by {\citet{mg14trs}}.

\begin{figure}[t]
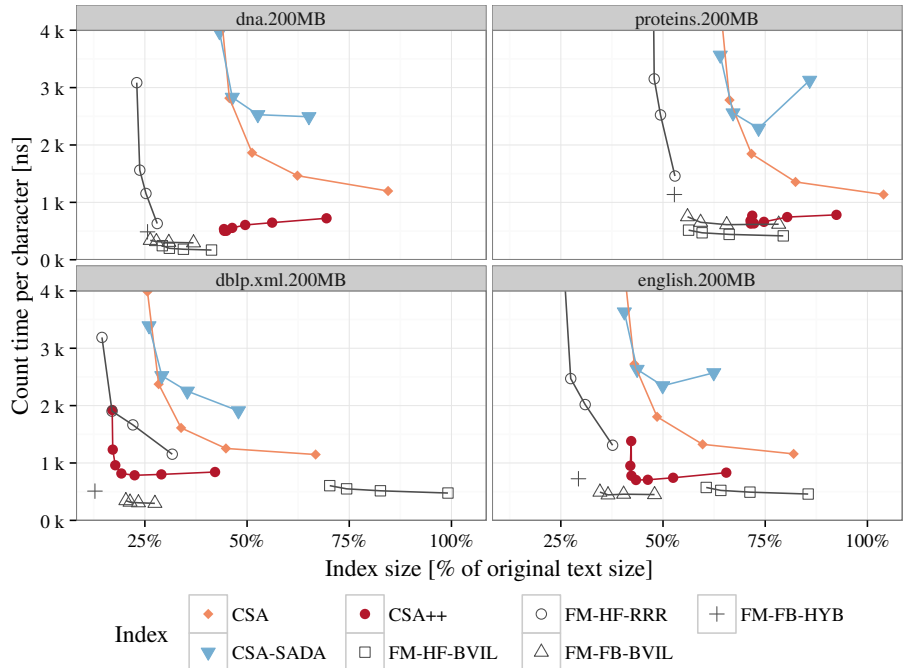

\centering

 \vspace*{-4mm}
\caption{Cost of indexed pattern search for a set of small- and
medium-alphabet Pizza\&Chili files, each $\mb{200}$.
The preferred zone is at the lower-left.
\label{fig-count-byte}}
\end{figure}

\myparagraph{Pattern Search, Small Alphabets}

Figure~\ref{fig-count-byte} depicts the relative performance of two
previous {\CSA} implementations, and a total four of FM-Index
options.
The method marked {\CSA} reflects the description of
{\cite{s-jalg03}}, as implemented in the SDSL library; it stores the
$\psi$ function using Elias $\gamma$ codes as a single stream of
gaps, with the disruptive elements at the start of each segment
represented as very large values rather than as negative gaps, and
with the samples stored uncompressed.
Method CSA-SADA is Sadakane's implementation of the same
mechanism.
The {\CSA++} is the approach described here.

We compare against two versions of each of two FM-Index approaches.
The first pair, prefixed FM-HF, use a Huffman-shaped wavelet tree
(WT) for the whole BWT {\citep{mn05cpm}}.
The first version of this approach represents the WT by an
uncompressed bitvector and a cache-friendly rank structure
(FM-HF-BVIL), and seeks to provide fast querying at the expense of
memory space; the second one uses entropy-compressed bitvectors
(FM-HF-RRR) to represent the WT, and is at the other extreme of the
space/speed tradeoff.
The second pair of FM-Indexes are based on fixed-block compression
boosting, prefixed FM-FB.
The BWT is partitioned into fixed-length blocks and a WT is created
for each block.
We use a recent implementation by {\citet{gkkpp2016dcc}}, and plug-in
an uncompressed bitvector and rank structure (FM-FB-BVIL), and a
hybrid bitvector (FM-FB-HYB) {\citep{kkp14dcc}}.
We did not have access to an implementation of another recent {\CSA}
proposal {\citep{aks15nsdi}}.

In Figure~\ref{fig-count-byte}, index size on the horizontal axis is
expressed as a percentage relative to the original text size, which
in the case of these four files, is always $\mb{200}$.
To measure search times, plotted on the vertical axis, the
corresponding query streams were executed in entirety to determine an
{\var{nocc}} count for each query, and then the overall execution
time for the stream was divided by the total number of query
characters, to obtain a computation time per query byte.
Where there is more than one point shown for a method, the blocksize
$k$ is the parameter being varied.
As can be seen from the four graphs in the figure, in general, the
best of the FM-Indexes tested were the two FM-FB variants, and they
also outperformed the two {\CSA} implementations.
The {\CSA++} outperforms both implementations of the earlier
{\CSA} approach on all four files.

\begin{figure}[t]
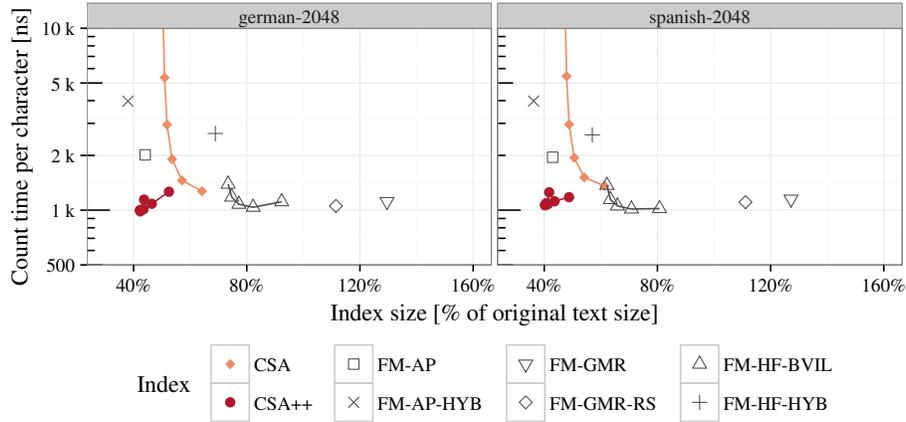

\centering

 \vspace*{-2mm}
\caption{Cost of indexed pattern search two $\gb{2}$ files of natural
language text, parsed in to word tokens.
The preferred zone is at the lower-left.
Note that the vertical axis is logarithmic.
\label{fig-count-int}}
\end{figure}

\myparagraph{Pattern Search, Large Alphabets}

Figure~\ref{fig-count-int} shows the same experiment, applied to the
large-alphabet natural-language texts.
A total of six FM-Index methods suited to large alphabets are
compared to the previous {\CSA} (the SDSL version) and the new
{\CSA++}: an alphabet partitioned (FM-AP) index {\citep{bgnn10isaac}}
which provides $O(\log\log\sigma)$ rank time, and a variant FM-AP-HYB
which uses a hybrid bitvector {\cite{kkp14dcc}}; two versions of
{\citeauthor{gmr06soda}}'s {\cite{gmr06soda}} rank structure (GMR-RS
and GMR); and again a huffman shaped WT using either a plain bitvector
(FM-HF-BVIL) or a hybrid bitvectors (FM-HF-HYB).
In this environment the {\CSA++} dominates all of the alternative
mechanisms, requiring either substantially less space, or offering
greatly improved query rates.
The careful attention paid to the representation of infrequent terms
is clearly beneficial.

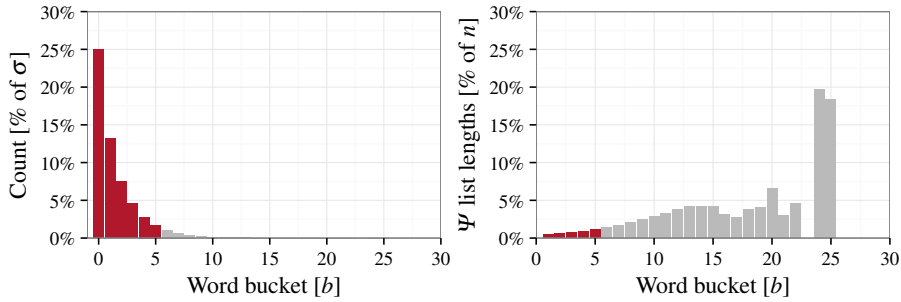
\begin{figure}[t]
\centering
\begin{tikzpicture}[x=1pt,y=1pt]
\definecolor{fillColor}{RGB}{255,255,255}
\path[use as bounding box,fill=fillColor,fill opacity=0.00] (0,0) rectangle (339.67,115.63);
\begin{scope}
\path[clip] (  0.00,  0.00) rectangle (169.83,115.63);
\definecolor{drawColor}{RGB}{255,255,255}
\definecolor{fillColor}{RGB}{255,255,255}

\path[draw=drawColor,line width= 0.6pt,line join=round,line cap=round,fill=fillColor] (  0.00,  0.00) rectangle (169.83,115.63);
\end{scope}
\begin{scope}
\path[clip] ( 32.05, 25.90) rectangle (165.83,111.63);
\definecolor{fillColor}{RGB}{255,255,255}

\path[fill=fillColor] ( 32.05, 25.90) rectangle (165.83,111.63);
\definecolor{drawColor}{gray}{0.98}

\path[draw=drawColor,line width= 0.6pt,line join=round] ( 32.05, 33.04) --
	(165.83, 33.04);

\path[draw=drawColor,line width= 0.6pt,line join=round] ( 32.05, 47.33) --
	(165.83, 47.33);

\path[draw=drawColor,line width= 0.6pt,line join=round] ( 32.05, 61.62) --
	(165.83, 61.62);

\path[draw=drawColor,line width= 0.6pt,line join=round] ( 32.05, 75.91) --
	(165.83, 75.91);

\path[draw=drawColor,line width= 0.6pt,line join=round] ( 32.05, 90.20) --
	(165.83, 90.20);

\path[draw=drawColor,line width= 0.6pt,line join=round] ( 32.05,104.49) --
	(165.83,104.49);

\path[draw=drawColor,line width= 0.6pt,line join=round] ( 47.15, 25.90) --
	( 47.15,111.63);

\path[draw=drawColor,line width= 0.6pt,line join=round] ( 68.73, 25.90) --
	( 68.73,111.63);

\path[draw=drawColor,line width= 0.6pt,line join=round] ( 90.31, 25.90) --
	( 90.31,111.63);

\path[draw=drawColor,line width= 0.6pt,line join=round] (111.89, 25.90) --
	(111.89,111.63);

\path[draw=drawColor,line width= 0.6pt,line join=round] (133.47, 25.90) --
	(133.47,111.63);

\path[draw=drawColor,line width= 0.6pt,line join=round] (155.05, 25.90) --
	(155.05,111.63);
\definecolor{drawColor}{gray}{0.90}

\path[draw=drawColor,line width= 0.2pt,line join=round] ( 32.05, 25.90) --
	(165.83, 25.90);

\path[draw=drawColor,line width= 0.2pt,line join=round] ( 32.05, 40.19) --
	(165.83, 40.19);

\path[draw=drawColor,line width= 0.2pt,line join=round] ( 32.05, 54.48) --
	(165.83, 54.48);

\path[draw=drawColor,line width= 0.2pt,line join=round] ( 32.05, 68.76) --
	(165.83, 68.76);

\path[draw=drawColor,line width= 0.2pt,line join=round] ( 32.05, 83.05) --
	(165.83, 83.05);

\path[draw=drawColor,line width= 0.2pt,line join=round] ( 32.05, 97.34) --
	(165.83, 97.34);

\path[draw=drawColor,line width= 0.2pt,line join=round] ( 32.05,111.63) --
	(165.83,111.63);

\path[draw=drawColor,line width= 0.2pt,line join=round] ( 36.37, 25.90) --
	( 36.37,111.63);

\path[draw=drawColor,line width= 0.2pt,line join=round] ( 57.94, 25.90) --
	( 57.94,111.63);

\path[draw=drawColor,line width= 0.2pt,line join=round] ( 79.52, 25.90) --
	( 79.52,111.63);

\path[draw=drawColor,line width= 0.2pt,line join=round] (101.10, 25.90) --
	(101.10,111.63);

\path[draw=drawColor,line width= 0.2pt,line join=round] (122.68, 25.90) --
	(122.68,111.63);

\path[draw=drawColor,line width= 0.2pt,line join=round] (144.26, 25.90) --
	(144.26,111.63);

\path[draw=drawColor,line width= 0.2pt,line join=round] (165.83, 25.90) --
	(165.83,111.63);
\definecolor{fillColor}{RGB}{178,24,43}

\path[fill=fillColor] ( 34.42, 25.90) rectangle ( 38.31, 97.24);

\path[fill=fillColor] ( 38.74, 25.90) rectangle ( 42.62, 63.59);

\path[fill=fillColor] ( 43.06, 25.90) rectangle ( 46.94, 47.43);

\path[fill=fillColor] ( 47.37, 25.90) rectangle ( 51.25, 38.88);

\path[fill=fillColor] ( 51.69, 25.90) rectangle ( 55.57, 33.85);

\path[fill=fillColor] ( 56.00, 25.90) rectangle ( 59.89, 30.85);
\definecolor{fillColor}{gray}{0.73}

\path[fill=fillColor] ( 60.32, 25.90) rectangle ( 64.20, 28.98);

\path[fill=fillColor] ( 64.63, 25.90) rectangle ( 68.52, 27.80);

\path[fill=fillColor] ( 68.95, 25.90) rectangle ( 72.83, 27.04);

\path[fill=fillColor] ( 73.26, 25.90) rectangle ( 77.15, 26.59);

\path[fill=fillColor] ( 77.58, 25.90) rectangle ( 81.46, 26.30);

\path[fill=fillColor] ( 81.90, 25.90) rectangle ( 85.78, 26.13);

\path[fill=fillColor] ( 86.21, 25.90) rectangle ( 90.10, 26.03);

\path[fill=fillColor] ( 90.53, 25.90) rectangle ( 94.41, 25.97);

\path[fill=fillColor] ( 94.84, 25.90) rectangle ( 98.73, 25.93);

\path[fill=fillColor] ( 99.16, 25.90) rectangle (103.04, 25.92);

\path[fill=fillColor] (103.47, 25.90) rectangle (107.36, 25.90);

\path[fill=fillColor] (107.79, 25.90) rectangle (111.67, 25.90);

\path[fill=fillColor] (112.11, 25.90) rectangle (115.99, 25.90);

\path[fill=fillColor] (116.42, 25.90) rectangle (120.30, 25.90);

\path[fill=fillColor] (120.74, 25.90) rectangle (124.62, 25.90);

\path[fill=fillColor] (125.05, 25.90) rectangle (128.94, 25.90);

\path[fill=fillColor] (129.37, 25.90) rectangle (133.25, 25.90);

\path[fill=fillColor] (138.00, 25.90) rectangle (141.88, 25.90);

\path[fill=fillColor] (142.31, 25.90) rectangle (146.20, 25.90);
\definecolor{drawColor}{gray}{0.50}

\path[draw=drawColor,line width= 0.6pt,line join=round,line cap=round] ( 32.05, 25.90) rectangle (165.83,111.63);
\end{scope}
\begin{scope}
\path[clip] (  0.00,  0.00) rectangle (339.67,115.63);
\definecolor{drawColor}{RGB}{0,0,0}

\node[text=drawColor,anchor=base east,inner sep=0pt, outer sep=0pt, scale=  0.80] at ( 27.55, 23.14) {0\%};

\node[text=drawColor,anchor=base east,inner sep=0pt, outer sep=0pt, scale=  0.80] at ( 27.55, 37.43) {5\%};

\node[text=drawColor,anchor=base east,inner sep=0pt, outer sep=0pt, scale=  0.80] at ( 27.55, 51.72) {10\%};

\node[text=drawColor,anchor=base east,inner sep=0pt, outer sep=0pt, scale=  0.80] at ( 27.55, 66.01) {15\%};

\node[text=drawColor,anchor=base east,inner sep=0pt, outer sep=0pt, scale=  0.80] at ( 27.55, 80.30) {20\%};

\node[text=drawColor,anchor=base east,inner sep=0pt, outer sep=0pt, scale=  0.80] at ( 27.55, 94.59) {25\%};

\node[text=drawColor,anchor=base east,inner sep=0pt, outer sep=0pt, scale=  0.80] at ( 27.55,108.88) {30\%};
\end{scope}
\begin{scope}
\path[clip] (  0.00,  0.00) rectangle (339.67,115.63);
\definecolor{drawColor}{RGB}{0,0,0}

\path[draw=drawColor,line width= 0.6pt,line join=round] ( 29.55, 25.90) --
	( 32.05, 25.90);

\path[draw=drawColor,line width= 0.6pt,line join=round] ( 29.55, 40.19) --
	( 32.05, 40.19);

\path[draw=drawColor,line width= 0.6pt,line join=round] ( 29.55, 54.48) --
	( 32.05, 54.48);

\path[draw=drawColor,line width= 0.6pt,line join=round] ( 29.55, 68.76) --
	( 32.05, 68.76);

\path[draw=drawColor,line width= 0.6pt,line join=round] ( 29.55, 83.05) --
	( 32.05, 83.05);

\path[draw=drawColor,line width= 0.6pt,line join=round] ( 29.55, 97.34) --
	( 32.05, 97.34);

\path[draw=drawColor,line width= 0.6pt,line join=round] ( 29.55,111.63) --
	( 32.05,111.63);
\end{scope}
\begin{scope}
\path[clip] (  0.00,  0.00) rectangle (339.67,115.63);
\definecolor{drawColor}{RGB}{0,0,0}

\path[draw=drawColor,line width= 0.6pt,line join=round] ( 36.37, 23.40) --
	( 36.37, 25.90);

\path[draw=drawColor,line width= 0.6pt,line join=round] ( 57.94, 23.40) --
	( 57.94, 25.90);

\path[draw=drawColor,line width= 0.6pt,line join=round] ( 79.52, 23.40) --
	( 79.52, 25.90);

\path[draw=drawColor,line width= 0.6pt,line join=round] (101.10, 23.40) --
	(101.10, 25.90);

\path[draw=drawColor,line width= 0.6pt,line join=round] (122.68, 23.40) --
	(122.68, 25.90);

\path[draw=drawColor,line width= 0.6pt,line join=round] (144.26, 23.40) --
	(144.26, 25.90);

\path[draw=drawColor,line width= 0.6pt,line join=round] (165.83, 23.40) --
	(165.83, 25.90);
\end{scope}
\begin{scope}
\path[clip] (  0.00,  0.00) rectangle (339.67,115.63);
\definecolor{drawColor}{RGB}{0,0,0}

\node[text=drawColor,anchor=base,inner sep=0pt, outer sep=0pt, scale=  0.80] at ( 36.37, 15.89) {0};

\node[text=drawColor,anchor=base,inner sep=0pt, outer sep=0pt, scale=  0.80] at ( 57.94, 15.89) {5};

\node[text=drawColor,anchor=base,inner sep=0pt, outer sep=0pt, scale=  0.80] at ( 79.52, 15.89) {10};

\node[text=drawColor,anchor=base,inner sep=0pt, outer sep=0pt, scale=  0.80] at (101.10, 15.89) {15};

\node[text=drawColor,anchor=base,inner sep=0pt, outer sep=0pt, scale=  0.80] at (122.68, 15.89) {20};

\node[text=drawColor,anchor=base,inner sep=0pt, outer sep=0pt, scale=  0.80] at (144.26, 15.89) {25};

\node[text=drawColor,anchor=base,inner sep=0pt, outer sep=0pt, scale=  0.80] at (165.83, 15.89) {30};
\end{scope}
\begin{scope}
\path[clip] (  0.00,  0.00) rectangle (339.67,115.63);
\definecolor{drawColor}{RGB}{0,0,0}

\node[text=drawColor,anchor=base,inner sep=0pt, outer sep=0pt, scale=  1.00] at ( 98.94,  5.00) {Word bucket [$b$]};
\end{scope}
\begin{scope}
\path[clip] (  0.00,  0.00) rectangle (339.67,115.63);
\definecolor{drawColor}{RGB}{0,0,0}

\node[text=drawColor,rotate= 90.00,anchor=base,inner sep=0pt, outer sep=0pt, scale=  1.00] at (  8.89, 68.76) {Count [\% of $\sigma$]};
\end{scope}
\begin{scope}
\path[clip] (169.83,  0.00) rectangle (339.67,115.63);
\definecolor{drawColor}{RGB}{255,255,255}
\definecolor{fillColor}{RGB}{255,255,255}

\path[draw=drawColor,line width= 0.6pt,line join=round,line cap=round,fill=fillColor] (169.83,  0.00) rectangle (339.67,115.63);
\end{scope}
\begin{scope}
\path[clip] (201.88, 25.90) rectangle (335.67,111.63);
\definecolor{fillColor}{RGB}{255,255,255}

\path[fill=fillColor] (201.88, 25.90) rectangle (335.67,111.63);
\definecolor{drawColor}{gray}{0.98}

\path[draw=drawColor,line width= 0.6pt,line join=round] (201.88, 33.04) --
	(335.67, 33.04);

\path[draw=drawColor,line width= 0.6pt,line join=round] (201.88, 47.33) --
	(335.67, 47.33);

\path[draw=drawColor,line width= 0.6pt,line join=round] (201.88, 61.62) --
	(335.67, 61.62);

\path[draw=drawColor,line width= 0.6pt,line join=round] (201.88, 75.91) --
	(335.67, 75.91);

\path[draw=drawColor,line width= 0.6pt,line join=round] (201.88, 90.20) --
	(335.67, 90.20);

\path[draw=drawColor,line width= 0.6pt,line join=round] (201.88,104.49) --
	(335.67,104.49);

\path[draw=drawColor,line width= 0.6pt,line join=round] (213.03, 25.90) --
	(213.03,111.63);

\path[draw=drawColor,line width= 0.6pt,line join=round] (235.33, 25.90) --
	(235.33,111.63);

\path[draw=drawColor,line width= 0.6pt,line join=round] (257.63, 25.90) --
	(257.63,111.63);

\path[draw=drawColor,line width= 0.6pt,line join=round] (279.93, 25.90) --
	(279.93,111.63);

\path[draw=drawColor,line width= 0.6pt,line join=round] (302.22, 25.90) --
	(302.22,111.63);

\path[draw=drawColor,line width= 0.6pt,line join=round] (324.52, 25.90) --
	(324.52,111.63);
\definecolor{drawColor}{gray}{0.90}

\path[draw=drawColor,line width= 0.2pt,line join=round] (201.88, 25.90) --
	(335.67, 25.90);

\path[draw=drawColor,line width= 0.2pt,line join=round] (201.88, 40.19) --
	(335.67, 40.19);

\path[draw=drawColor,line width= 0.2pt,line join=round] (201.88, 54.48) --
	(335.67, 54.48);

\path[draw=drawColor,line width= 0.2pt,line join=round] (201.88, 68.76) --
	(335.67, 68.76);

\path[draw=drawColor,line width= 0.2pt,line join=round] (201.88, 83.05) --
	(335.67, 83.05);

\path[draw=drawColor,line width= 0.2pt,line join=round] (201.88, 97.34) --
	(335.67, 97.34);

\path[draw=drawColor,line width= 0.2pt,line join=round] (201.88,111.63) --
	(335.67,111.63);

\path[draw=drawColor,line width= 0.2pt,line join=round] (201.88, 25.90) --
	(201.88,111.63);

\path[draw=drawColor,line width= 0.2pt,line join=round] (224.18, 25.90) --
	(224.18,111.63);

\path[draw=drawColor,line width= 0.2pt,line join=round] (246.48, 25.90) --
	(246.48,111.63);

\path[draw=drawColor,line width= 0.2pt,line join=round] (268.78, 25.90) --
	(268.78,111.63);

\path[draw=drawColor,line width= 0.2pt,line join=round] (291.07, 25.90) --
	(291.07,111.63);

\path[draw=drawColor,line width= 0.2pt,line join=round] (313.37, 25.90) --
	(313.37,111.63);

\path[draw=drawColor,line width= 0.2pt,line join=round] (335.67, 25.90) --
	(335.67,111.63);
\definecolor{fillColor}{RGB}{178,24,43}

\path[fill=fillColor] (204.34, 25.90) rectangle (208.35, 27.25);

\path[fill=fillColor] (208.80, 25.90) rectangle (212.81, 27.52);

\path[fill=fillColor] (213.26, 25.90) rectangle (217.27, 27.90);

\path[fill=fillColor] (217.72, 25.90) rectangle (221.73, 28.38);

\path[fill=fillColor] (222.18, 25.90) rectangle (226.19, 29.02);
\definecolor{fillColor}{gray}{0.73}

\path[fill=fillColor] (226.63, 25.90) rectangle (230.65, 29.78);

\path[fill=fillColor] (231.09, 25.90) rectangle (235.11, 30.71);

\path[fill=fillColor] (235.55, 25.90) rectangle (239.57, 31.69);

\path[fill=fillColor] (240.01, 25.90) rectangle (244.03, 32.87);

\path[fill=fillColor] (244.47, 25.90) rectangle (248.49, 34.05);

\path[fill=fillColor] (248.93, 25.90) rectangle (252.95, 35.19);

\path[fill=fillColor] (253.39, 25.90) rectangle (257.41, 36.86);

\path[fill=fillColor] (257.85, 25.90) rectangle (261.86, 37.90);

\path[fill=fillColor] (262.31, 25.90) rectangle (266.32, 37.72);

\path[fill=fillColor] (266.77, 25.90) rectangle (270.78, 37.77);

\path[fill=fillColor] (271.23, 25.90) rectangle (275.24, 34.75);

\path[fill=fillColor] (275.69, 25.90) rectangle (279.70, 33.61);

\path[fill=fillColor] (280.15, 25.90) rectangle (284.16, 36.57);

\path[fill=fillColor] (284.61, 25.90) rectangle (288.62, 37.39);

\path[fill=fillColor] (289.07, 25.90) rectangle (293.08, 44.65);

\path[fill=fillColor] (293.53, 25.90) rectangle (297.54, 34.36);

\path[fill=fillColor] (297.99, 25.90) rectangle (302.00, 38.84);

\path[fill=fillColor] (306.91, 25.90) rectangle (310.92, 82.07);

\path[fill=fillColor] (311.36, 25.90) rectangle (315.38, 78.45);
\definecolor{drawColor}{gray}{0.50}

\path[draw=drawColor,line width= 0.6pt,line join=round,line cap=round] (201.88, 25.90) rectangle (335.67,111.63);
\end{scope}
\begin{scope}
\path[clip] (  0.00,  0.00) rectangle (339.67,115.63);
\definecolor{drawColor}{RGB}{0,0,0}

\node[text=drawColor,anchor=base east,inner sep=0pt, outer sep=0pt, scale=  0.80] at (197.38, 23.14) {0\%};

\node[text=drawColor,anchor=base east,inner sep=0pt, outer sep=0pt, scale=  0.80] at (197.38, 37.43) {5\%};

\node[text=drawColor,anchor=base east,inner sep=0pt, outer sep=0pt, scale=  0.80] at (197.38, 51.72) {10\%};

\node[text=drawColor,anchor=base east,inner sep=0pt, outer sep=0pt, scale=  0.80] at (197.38, 66.01) {15\%};

\node[text=drawColor,anchor=base east,inner sep=0pt, outer sep=0pt, scale=  0.80] at (197.38, 80.30) {20\%};

\node[text=drawColor,anchor=base east,inner sep=0pt, outer sep=0pt, scale=  0.80] at (197.38, 94.59) {25\%};

\node[text=drawColor,anchor=base east,inner sep=0pt, outer sep=0pt, scale=  0.80] at (197.38,108.88) {30\%};
\end{scope}
\begin{scope}
\path[clip] (  0.00,  0.00) rectangle (339.67,115.63);
\definecolor{drawColor}{RGB}{0,0,0}

\path[draw=drawColor,line width= 0.6pt,line join=round] (199.38, 25.90) --
	(201.88, 25.90);

\path[draw=drawColor,line width= 0.6pt,line join=round] (199.38, 40.19) --
	(201.88, 40.19);

\path[draw=drawColor,line width= 0.6pt,line join=round] (199.38, 54.48) --
	(201.88, 54.48);

\path[draw=drawColor,line width= 0.6pt,line join=round] (199.38, 68.76) --
	(201.88, 68.76);

\path[draw=drawColor,line width= 0.6pt,line join=round] (199.38, 83.05) --
	(201.88, 83.05);

\path[draw=drawColor,line width= 0.6pt,line join=round] (199.38, 97.34) --
	(201.88, 97.34);

\path[draw=drawColor,line width= 0.6pt,line join=round] (199.38,111.63) --
	(201.88,111.63);
\end{scope}
\begin{scope}
\path[clip] (  0.00,  0.00) rectangle (339.67,115.63);
\definecolor{drawColor}{RGB}{0,0,0}

\path[draw=drawColor,line width= 0.6pt,line join=round] (201.88, 23.40) --
	(201.88, 25.90);

\path[draw=drawColor,line width= 0.6pt,line join=round] (224.18, 23.40) --
	(224.18, 25.90);

\path[draw=drawColor,line width= 0.6pt,line join=round] (246.48, 23.40) --
	(246.48, 25.90);

\path[draw=drawColor,line width= 0.6pt,line join=round] (268.78, 23.40) --
	(268.78, 25.90);

\path[draw=drawColor,line width= 0.6pt,line join=round] (291.07, 23.40) --
	(291.07, 25.90);

\path[draw=drawColor,line width= 0.6pt,line join=round] (313.37, 23.40) --
	(313.37, 25.90);

\path[draw=drawColor,line width= 0.6pt,line join=round] (335.67, 23.40) --
	(335.67, 25.90);
\end{scope}
\begin{scope}
\path[clip] (  0.00,  0.00) rectangle (339.67,115.63);
\definecolor{drawColor}{RGB}{0,0,0}

\node[text=drawColor,anchor=base,inner sep=0pt, outer sep=0pt, scale=  0.80] at (201.88, 15.89) {0};

\node[text=drawColor,anchor=base,inner sep=0pt, outer sep=0pt, scale=  0.80] at (224.18, 15.89) {5};

\node[text=drawColor,anchor=base,inner sep=0pt, outer sep=0pt, scale=  0.80] at (246.48, 15.89) {10};

\node[text=drawColor,anchor=base,inner sep=0pt, outer sep=0pt, scale=  0.80] at (268.78, 15.89) {15};

\node[text=drawColor,anchor=base,inner sep=0pt, outer sep=0pt, scale=  0.80] at (291.07, 15.89) {20};

\node[text=drawColor,anchor=base,inner sep=0pt, outer sep=0pt, scale=  0.80] at (313.37, 15.89) {25};

\node[text=drawColor,anchor=base,inner sep=0pt, outer sep=0pt, scale=  0.80] at (335.67, 15.89) {30};
\end{scope}
\begin{scope}
\path[clip] (  0.00,  0.00) rectangle (339.67,115.63);
\definecolor{drawColor}{RGB}{0,0,0}

\node[text=drawColor,anchor=base,inner sep=0pt, outer sep=0pt, scale=  1.00] at (268.78,  5.00) {Word bucket [$b$]};
\end{scope}
\begin{scope}
\path[clip] (  0.00,  0.00) rectangle (339.67,115.63);
\definecolor{drawColor}{RGB}{0,0,0}

\node[text=drawColor,rotate= 90.00,anchor=base,inner sep=0pt, outer sep=0pt, scale=  1.00] at (178.72, 68.76) {$\Psi$ list lengths [\% of $n$]};
\end{scope}
\end{tikzpicture}
 \vspace*{-2mm}
\caption{Frequency statistics for german-2048.
Bucket $b$ contains words which occur between $2^b$ and $2^{b+1}-1$
times, with bucket $b=0$ covering words that occur exactly once.
The left pane shows the percentage of $\sigma$ accounted for by each
bucket, and the right pane the percentage of $n$.
The low-frequency part of each distribution (less than $k=128$ word
occurrences) is marked in red.
\label{fig-context-stats}}
\end{figure}

Figure~\ref{fig-context-stats} helps explain the situation.
The great majority of the symbols in $\Sigma$ occur fewer than
$k=128$ times; indeed, $25$\% of them appear only once.
Reducing the per-term overhead is thus very important.
However, as is shown in the right pane, those terms are a small
percentage of the $\psi$ array, and storing them in binary is not
detrimental to overall performance.

\begin{table}[t]
\centering
\newcommand{\tabent}[1]{\makebox[11mm][c]{#1}}
\begin{tabular}{l@{~~}l p{0.5em} cc p{0.5em} cc p{0.5em} cc}
\toprule
\multirow{2}{*}{Method}
& \multirow{2}{*}{Component}
	&& \multicolumn{2}{c}{DNA (\mb{200})}
			&& \multicolumn{2}{c}{XML (\mb{200})}
					&& \multicolumn{2}{c}{German (\gb{2})}
\\
\cmidrule{4-5}\cmidrule{7-8}\cmidrule{10-11}
	&&& \tabent{\%\,{\raisebox{1.3pt}{$\psi$}}}
		& \tabent{MiB}
			&& \tabent{\%\,{\raisebox{1.3pt}{$\psi$}}}
				& \tabent{MiB}
					&& \tabent{\%\,{\raisebox{1.3pt}{$\psi$}}}
						& \tabent{MiB}
\\
\midrule
CSA++
& Samples
	&& --
		&\D\D2.3
			&& --
				&\D\D3.1
					&& --
						&\D36.8
\\
& NIL-blocks
	&&\D\D0.2
		&\D\D0.0
			&&\D62.0
				&\D\D0.0
					&&\D15.7
						&\D\D0.0
\\
& BV-coded blocks
	&&\D78.0
		&\D61.7
			&&\D10.8
				&\D\D5.9
					&&\D16.4
						&\D32.1
\\
& RL-coded blocks
	&&\D\D2.5
		&\D\D0.2
			&&\D14.4
				&\D\D4.3
					&&\D\D3.8
						&\D13.6
\\
& EF-coded blocks
	&&\D19.3
		&\D22.6
			&&\D12.7
				&\D20.4
					&&\D59.6
						&626.3
\\
& Binary values
	&&\D\D0.0
		&\D\D0.0
			&&\D\D0.0
				&\D\D0.0
					&&\D\D4.4
						&115.5
\\
& Other structures
	&& --
		&\D\D5.9
			&& --
				&\D\D4.8
					&& --
						&\D43.0
\\
& \emph{Total space}
	&& --
		&\D92.7
			&& --
				&\D38.5
					&& --
						&867.2
\\[0.5ex]
\cmidrule{1-11}
\multicolumn{2}{l}{CSA}
	&& --
		&\D91.4
			&& --
				&\D56.7
					&& --
						& 1061
\\
\multicolumn{2}{l}{FM-FB-HYB}
	&& --
		&\D51.3
			&& --
				&\D25.6
					&& --
						& --
\\
\multicolumn{2}{l}{FM-HF-HYB}
	&& --
		&\D51.8
			&& --
				&\D32.4
					&& --
						&1411
\\
\multicolumn{2}{l}{FM-AP}
	&& --
		& --
			&& --
				& --
					&& --
						&903.3
\\
\multicolumn{2}{l}{FM-AP-HYB}
	&& --
		& --
			&& --
				& --
					&& --
						&778.5
\\
\bottomrule
\end{tabular}
 \caption{Comparing the space costs of different pattern search
indexes, using a blocksize of $k=128$ throughout.
The methods listed in the lower part of the table are from the
SDSL library.
Note that not all of the methods are applicable to all of the files.
\label{tbl-bigtable}}
\end{table}

\myparagraph{Detailed Space Breakdown}

Table~\ref{tbl-bigtable} provide details of the space required by
various components of the improved {\CSA}, for a small-alphabet file,
a mid-alphabet file, and a large-alphabet file.
The two columns associated with each of the three files show the
space required by the named component, preceded by, where
appropriate, the fraction of the values in $\psi$ that are handled
via that option.
The EF-coded samples require around $1$--$2$\% of the original space;
and various other access structures, including the wavelet tree for
low-$n_c$ symbols, require a further $2$--$3$\%.
The four different block types play different roles across the three
files.
For the DNA data, the great majority of $\psi$ values are included in
BV blocks; for the XML data, the emphasis is on NIL blocks; and for
the word-based large-alphabet data it is EF blocks that dominate.
In the latter case, a small but important fraction of the $\psi$
values are coded in plain binary, as shown above in
Figure~\ref{fig-context-stats}.
The effect of this alphabet partitioning is better compression for
the EF-coded values, which on this file are the dominant type;
confirming that this option is an important component of the
large-alphabet situations handled so well by the {\CSA++}.
Table~\ref{tbl-bigtable} also lists the space needed by several other
compressed pattern search structures, to provide further context for
these results.

\begin{wraptable}{r}{55mm}
\vspace*{-8mm}
\begin{tabular}{c c cc c cc}
\toprule
\multirow{2}{*}{Blocksize}
	&& \multicolumn{2}{c}{{Random}}
		&& \multicolumn{2}{c}{{RLZ factors}}
\\
\cmidrule{3-4}\cmidrule{6-7}
	&& \CSA & \CSA++
		&& \CSA & \CSA++
\\
\midrule
$k=64$
	&& 1.84 & 0.74
		&& 1.68 & 0.56
\\
$k=128$
	&& 2.89 & 0.76
		&& 2.73 & 0.59
\\
$k=256$
	&& 5.10 & 0.88
		&& 4.90 & 0.73
\\
\bottomrule
\end{tabular}

 \vspace*{-2mm}
\caption{Per-character time in microseconds for RLZ factorization,
compared to $23$-character random patterns.
\vspace*{-6mm}
\label{tbl-rlzspeeds}}
\end{wraptable}

\myparagraph{Case Study, Text Factorization}

The {\emph{Relative Lempel-Ziv}} (RLZ) compression mechanism
represents a string $\STR$ as a sequence of {\emph{factors}} from
a dictionary $D$, see {\citet{pmnw15airs}} for a description
and experimental results.
To greedily determine longest factors using a {\CSA}, we take
$\TEXT=D^{\mbox{\scriptsize\sf{r}}}$, the reverse of $D$, and build a
compressed index.
The string is then processed against $\TEXT$ taking symbols from
$\STR$ in left-to-right order, and performing a backward search in
$\TEXT$; if a prefix of length $p$ from $\STR$ is sufficient to
ensure that the $\spep$ range becomes empty, then the next factor
emitted is of length $p-1$.
That is, the factorization process can be regarded as applying
variable-length patterns to a text $\TEXT$, with each pattern being
as short as possible without appearing in $\TEXT$.
To carry out an application-driven experiment, we took the $\gb{64}$
prefix of the GOV2 document collection used by
{\citeauthor{pmnw15airs}}, and built a set of patterns, each of which
is one factor, plus the next character from $\STR$.
The first $1{,}901{,}131{,}365$ patterns from that set, representing
$\gb{4}$ of text, were used as queries.
The average factor length was $23.6$ characters, with $\var{nocc}=0$
in $\TEXT$ in all cases.
We then applied those patterns to an $\mb{256}$ dictionary $D$
constructed from the whole $\gb{64}$, to compute the per-character
cost of performing the specified searches, and compared against the
per-character cost associated with search for randomly selected
patterns.
Table~\ref{tbl-rlzspeeds} shows the cost of backward search step in
both scenarios and confirms both that {\CSA++} significantly
outperforms {\CSA}, and also that for {\emph{count}} queries, random
strings are a reasonable experimental methodology.

\begin{wraptable}{r}{55mm}
\vspace*{-8mm}
\begin{tabular}{c c cc c cc}
\toprule
\multirow{2}{*}{Blocksize}
	&& \multicolumn{2}{c}{Random}
		&& \multicolumn{2}{c}{NL search}
\\
\cmidrule{3-4}\cmidrule{6-7}
	&& \CSA & \CSA++
		&& \CSA & \CSA++
\\
\midrule
$k=64$
	&& 1.86 & 1.05
		&& 1.67 & 0.63
\\
$k=128$
	&& 2.90 & 0.99
		&& 2.98 & 0.62
\\
$k=256$
	&& 5.50 & 0.99
		&& 5.99 & 0.63
\\
\bottomrule
\end{tabular}

 \vspace*{-2mm}
\caption{Per-word time in microseconds for phrase search,
compared to $4$-word random patterns.
\vspace*{-6mm}
\label{tbl-nlspeeds}}
\end{wraptable}

\myparagraph{Case Study, Language Modeling}

A common operation on natural language files is to identify
informative phrases as sentences are parsed {\cite{sphc-emnlp15}}.
We built variable-length queries for the file german-2048, and
measured the per-symbol processing time, comparing actual-use
queries and randomly-selected-string queries for {\CSA} and
{\CSA++}.
In total $1{,}521{,}869$ queries of average length $3.4$ words were
extracted from the machine translation process described by
{\citeauthor{sphc-emnlp15}}, corresponding to $40{,}000$ sentences
randomly selected from the German part of Common Crawl.
Table~\ref{tbl-nlspeeds} shows the cost of those {\emph{count}}
queries over the german-2048 file.
The results again align with the performance of pattern searches for
random queries extracted from the text, as was shown in
Figure~\ref{fig-count-int}.
Note in particular that {\CSA++} performance is largely unaffected by
$k$, whereas the performance of {\CSA} substantially decreases as $k$
increases.
As pattern search is a major part of the cost of the machine
translation process described by {\citet{sphc-emnlp15}}, utilizing
{\CSA++} leads to a significant speedup in practical performance.

\section{Conclusion}
\label{sec-conclusion}

We have described several enhancement's to Sadakane's {\CSA}, and
have demonstrated improvements both in terms of compression
effectiveness, and also in terms of query throughput for
{\emph{count}} queries, especially for large-alphabet applications.
If {\emph{locate}} queries are also required, all of the structures
explored here must be augmented with {\SUFF} samples, to allow
$\spep$ ranges to be converted to offsets in $\TEXT$; as future work,
we plan to investigate space-speed tradeoffs in that regard as well.

\myparagraph{Acknowledgment}
This work was supported under Australian Research Council's Discovery
Projects funding scheme (project number DP140103256).

\renewcommand{\bibfont}{\small}
\setlength\bibitemsep{2.0pt}
\printbibliography

\newpage
\section*{Appendix}

\subsection*{Details of Implementations}

Table~\ref{tbl-sdsl-details} provides details of the methods compared
in Section~\ref{sec-experiments}.
The CSA-SADA results were obtained by executing code authored
by Kunihiko Sadakane, available from the Pizza\&Chili web site.

\begin{table}[tbh!]
\begin{center}
\begin{tabular}{lp{1em}l}
\toprule
Abbreviation
	&& \multicolumn{1}{c}{Composition}
\\
\midrule
CSA
	&& \verb@csa_sada<enc_vector<coder::elias_gamma,@$s_{\Psi}$\verb@>,@\\
    && \verb@         1<<20,1<<20,sa_order_sa_sampling<>,isa_sampling<>>@
\\
FM-HF-BVIL
	&& \verb@csa_wt<wt_huff<bit_vector_il<@$\var{bs}$\verb@>>,1<<20,1<<20>@
\\
FM-HF-HYB
	&& \verb@csa_wt<wt_huff<hyb_vector<>>,1<<20,1<<20>@
\\
FM-HF-RRR
	&& \verb@csa_wt<wt_huff<rrr_vector<@$b$\verb@>>,1<<20,1<<20>@
\\
FM-AP
	&& \verb@csa_wt_int<wt_ap< wt_huff<bit_vector,rank_support_v5<1>,@\\
    && \verb@           select_support_scan<1>,select_support_scan<0>>,@\\
    && \verb@           wm_int<bit_vector,rank_support_v5<1>,@\\
    && \verb@           select_support_scan<1>,select_support_scan<0>>>,@\\
    && \verb@           1<<20,1<<20>@
\\
FM-AP-HYB
	&& \verb@csa_wt_int<wt_ap< wt_huff<hyb_vector<>>,@\\
    && \verb@           wm_int<hyb_vector<>> >,1<<20,1<<20>@
\\
FM-GMR
	&& \verb@csa_wt_int<wt_gmr<>,1<<20,1<<20>@
\\
FM-GMR-RS
	&& \verb@csa_wt_int<wt_gmr_rs<>,1<<20,1<<20>@
\\\hline
FM-FB-BVIL
	&& \verb@csa_wt<wt_fbb<bit_vector_il<@$\var{bs}$\verb@>>,1<<20,1<<20>@
\\
FM-FB-HYB
	&& \verb@csa_wt<wt_fbb<hyb_vector<>>,1<<20,1<<20>@
\\
CSA++
	&& \verb@csa_sada2<hyb_sd_vector<@$s$\verb@>,1<<20,1<<20,@\\
    && \verb@          sa_order_sa_sampling<>,isa_sampling<>>@
\\

\bottomrule
\end{tabular}
 \end{center}
\caption{SDSL descriptions of methods used in experiments.
Sampling parameters $b\in \{15,31,63,127\}$, $\var{bs}\in\{128,256,512,1024\}$,
$s\in\{16,32,64,128,256,512,1024\}$, and $s_{\Psi}\in \{16,32,64,128,512,1024\}$
were varied in the experiments
to get different time-space trade-offs. The last three class
definitions are available in the \texttt{hyb\_sd\_vector}
branch of the library.
\label{tbl-sdsl-details}}
\end{table}

\end{document}